\documentclass[useAMS,usenatbib]{mn2e}

\usepackage{times}
\usepackage{epsf}
\usepackage[dvips]{epsfig}
\usepackage{amsmath}

\newcommand{\bd}{\begin{displaymath}}
\newcommand{\ed}{\end{displaymath}}
\def\ZZ#1{$\scriptstyle #1$}
\newcommand{\be}{\begin{equation}}
\newcommand{\ee}{\end{equation}}
\newcommand{\bea}{\begin{eqnarray}}
\newcommand{\eea}{\end{eqnarray}}

\title[Post-Newtonian simulations]{Post-Newtonian $N$-body simulations}
\author[Sverre J. Aarseth]{Sverre J. Aarseth \thanks{E-mail:
sverre@ast.cam.ac.uk}\\
Institute of Astronomy, Madingley Road, Cambridge CB3 0HA}

\begin{document}
\date{Accepted 2007 January XX. Received 2007 January YY; in original form 2007 January 12}

\pagerange{\pageref{firstpage}--\pageref{lastpage}} \pubyear{2007}

\maketitle

\label{firstpage}

\begin{abstract}
We report on the first fully consistent conventional cluster simulation
which includes terms up to ${\rm post}^{5/2}$ Newtonian in the potential
of the massive body.
Numerical problems for treating extremely energetic binaries orbiting a
single massive object are circumvented by employing the special
``wheel-spoke'' regularization method of Zare (1974) which has not been
used in large-$N$ simulations before.
Idealized models containing $N = 1 \times 10^5$ particles of mass
$1~{\rm M}_{\odot}$ with a central black hole of $300~{\rm M}_{\odot}$
have been studied on GRAPE-type computers.
An initial half-mass radius of $r_{\rm h} \simeq 0.1$~pc is sufficiently
small to yield examples of relativistic coalescence.
This is achieved by significant binary shrinkage within a density cusp
environment, followed by the generation of extremely high eccentricities
which are induced by Kozai (1962) cycles and/or resonant relaxation.
More realistic models with white dwarfs and ten times larger half-mass
radii also show evidence of GR effects before disruption.
Experimentation with the post-Newtonian terms suggests that reducing
the time-scales for activating the different orders progressively may be
justified for obtaining qualitatively correct solutions without aiming
for precise predictions of the final gravitational radiation wave form.
The results obtained suggest that the standard loss-cone arguments 
underestimate the swallowing rate in globular clusters containing a
central black hole.
\end{abstract}


\section{INTRODUCTION}

$N$-body simulations of stellar systems containing one or more massive
bodies have become popular with a number of studies
(Milosavlevi\'c \& Merritt 2001, Baumgardt, Makino \& Ebisuzaki 2004,
Baumgardt, Hopman, Portegies Zwart \& Makino 2006).
This type of work has been inspired by recent determinations of massive
black holes (BHs) which reside in galaxies and intermediate mass
black holes (IMBHs) have also been considered, particularly in relation
to globular clusters.
Although the presence of BHs has motivated such work, few papers so
far have included the relativistic effects that would be required
for an appropriate treatment of deviations from Newtonian motions.
One reason for this situation is the numerical challenge of dealing
with point-mass dynamics, both as regards the large range in time-scales
as well as nearly singular interactions involving superhard binaries.
Regularization methods have traditionally been used to overcome problems
of this kind.
However, existing codes are not tailor-made for dealing with large
mass ratios and some attempts to employ two-body (Milosavlevi\'c \&
Merritt 2001) and chain regularization (Szell, Merritt \& Mikkola 2005,
Amaro-Seoane \& Freitag 2006) only succeeded in obtaining modest shrinkage
of the dominant binary.

Binary shrinkage by itself is unlikely to be be sufficient in achieving
relativistic stages for all but the most extreme initial conditions.
Thus, given two BH binary mass components $m_{\rm BH} = N^{1/2} \bar m$,
where $\bar m$ is the average mass of $N$ field stars, the final semi-major
axis for absorbing half the total energy would be $4 r_{\rm h}/N$ in
terms of the half-mass radius $r_{\rm h}$.
Such evolution is well within practical reach for $N = 10^5$ with present
methods.
However, this would fall well short of the required shrinkage for
relativistic conditions using conventional length-scales.
The question therefore arises whether large eccentricities may be generated,
thereby reducing the relativistic time-scale significantly by the classical
factor $(1 - e^2)^{7/2}$.

Results obtained so far have yielded conflicting information about the
eccentricity evolution but clues are emerging that this may well depend
on initial conditions.
An early simulation based on two merging clusters, each containing a massive
single object inside a cusp-like system of $N \simeq 1 \times 10^{5}$
particles, produced sufficiently large values ($e_{\rm max} > 0.995$) for
the GR coalescence condition (Aarseth 2003b).
This work employed the time-transformed leapfrog method ({\ZZ {TTL}}) which was
specially designed to deal with a massive binary BH (Mikkola \& Aarseth 2002)
and included the three first post-Newtonian terms.
Large eccentricities were also reproduced recently by placing two massive
objects inside a single rotating cluster (Berczik et al.~2006).
A new study of three massive BHs by direct integration (Iwasawa, Funato
\& Makino 2006) did achieve coalescence by including relativistic energy
loss, albeit for large mass ratios and short times.

Although the binary BH problem is rightly receiving considerable attention,
the simpler case of just one massive object is still waiting to be explored
fully.
So far only one major investigation appears to have been directed towards
this goal (Preto, Merritt \& Spurzem 2004).
In this study, the attention was focused on the growth of a density cusp
around a massive central object for comparison with Fokker--Planck solutions.
In spite of using chain regularization (Mikkola \& Aarseth 1993), the maximum
growth in $1/a_{\rm BH}$ was relatively modest, mainly due to the mass ratio
of about 1000 exceeding somewhat the above-mentioned factor of
$N^{1/2}$ for $N$ in the range $1 - 2.5 \times 10^5$.

On the theoretical side, some efforts have been made to look for
enhanced relaxation in connection with the loss-cone problem.
The interesting concept of resonant relaxation (Rauch \& Tremaine 1996)
has yet to be tested in large-scale simulations.
More recent work (Hopman \& Alexander 2006) has hinted at a connection
with the Kozai mechanism.
On the basis of Fokker--Planck simulations it is suggested that coherent
torques change the angular momenta and rotate the orbital inclinations,
giving an increased tidal disruption rate for the central region.

The present investigation concentrates on the evolution of the central
subsystem using a powerful new implementation.
The basic method is a generalization of three-body regularization
(Aarseth \& Zare 1974) which is of the same vintage (Zare 1974) but apart
from one application to four-body scattering (Alexander 1986) has apparently
remained dormant.
Because of the analogy with spokes on a cart-wheel being connected to the
hub, it has been termed ``wheel-spoke regularization'' (Aarseth 2003a) since
each spoke represents mutual interactions with respect to the central hub
or reference body.
Experience with the three-body regularization code suggests that the
extension to arbitrary memberships would make an ideal tool for studying
the single BH problem, as was already suggested when the phrase was coined.
Such a scheme has the advantage of the massive object acting as a permanent
reference body, thereby allowing considerable simplifications compared to
the standard chain algorithm.
At the same time, post-Newtonian terms are included as perturbations in
anticipation of favourable developments for which very high eccentricities
would be needed.
A full post-Newtonian treatment has already been tried in a different
context (Kupi, Amaro-Seoane \& Spurzem 2006) for an extremely compact system
of 1000 point-mass particles with rms velocity about one percent of the
speed of light.
The aim here was to study run-away evolution by capture and mergers due to
gravitational radiation using two-body regularization, as was first done
a long time ago with the code {\ZZ {NBODY5}} (Lee 1993) without the
precession terms.

This paper begins by outlining the basic method and its implementation
in an $N$-body system, followed by a description of the initial conditions
for an equilibrium distribution.
We report on the results of several similar idealized models which
illustrate the capabilities of the method.
Realistic models which include finite-size effects involving white dwarfs
are also considered.
Some features of the post-Newtonian implementation are described, with
emphasis on the experimental nature of the scheme.
Finally a brief discussion is given, together with some suggestions for
the future.

\section{WHEEL-SPOKE IMPLEMENTATION}

We begin by defining a subsystem of $n$ single particles of mass $m_i$ and
one dominant body denoted by $i = 0$ with mass $m_0$.
The initial coordinates and momenta in the local centre of mass are
given by ${\tilde {\bf q}}_i, \,{\tilde {\bf p}}_i$ for $ i = 0,\ldots,n$.
Let us take $\bf p_0 = 0$ for the reference body and introduce relative
coordinates $\bf q_i = {\tilde {\bf q}}_i - {\tilde {\bf q}}_0$ with
respect to $m_0$.
The Hamiltonian then takes the form
\be
H \,=\, \sum_{i=1}^{n}\frac{{\bf p}_i^2}{2 \mu_i} +
    \frac{1}{m_0} \sum_{i<j}^{n} {\bf p}_i^{\rm T} \cdot {\bf p}_j -
 m_0 \sum_{i=1}^{n} \frac {m_i}{R_i}
 - \sum_{i<j}^{n}\frac{m_i m_j}{R_{ij}} \,,
\ee
with $\mu_i = m_i m_0 /(m_i + m_0)$ and $R_i = \vert \bf q_i \vert$.

Canonical variables $\bf Q_i, \, \bf P_i$ with a separable generating
function are now introduced for each two-body pair $m_i, m_0$ by
(Zare 1974)
\be
 W({\bf p}_i,{\bf Q}_i) =
 \sum_{i=1}^{n}{\bf p}_i^{\rm T} \cdot {\bf f}_i({\bf Q}_i) \,,
\ee
where ${\bf f}_i({\bf Q}_i)$ connects physical and regularized
coordinates for spoke index $i$.
The corresponding regularized momenta take the form
\be
{\bf P}_i = {\bf A}_i {\bf p}_i \,, ~~~(i=1,\ldots,n) \,,
\ee
where ${\bf A}_i$ is {\it twice} the transpose $4 \times 4$
Levi--Civita matrix.
Inverse transformations yield the relative coordinates and momenta
\be
 {\bf q}_i = \frac{1}{2} \,{\bf A}_i^{\rm T} {\bf Q}_i \,,~~~~
 {\bf p}_i = \frac{1}{4}\, \frac {{\bf A}_i^{\rm T} {\bf P}_i} { R_i} \,,
\ee
from which the final physical variables are readily determined.
Regularized equations of motion for each spoke interaction are finally
obtained by introducing a time transformation.
In accordance with standard practice in multiple regularization
(Alexander 1986, Mikkola \& Aarseth 1993) we adopt the inverse Lagrangian,
$t' = 1/L$, which has proved effective.
Differentiation of the regularized Hamiltonian $\Gamma^{\ast} = t' (H - E_0)$
(where $E_0$ is the total energy) with respect to $\bf Q$ and $\bf P$ then
yields the equations of motion to be integrated.

The implementation of the wheel-spoke scheme into a general $N$-body code
requires many special procedures.
Quite a few of these can be taken over from other large simulation codes
(Aarseth 2003a) when due allowance is made for differences in the data
structure.
In the following we summarize some of the most relevant features which
have much in common with chain regularization.

To begin with, a suitable subsystem must first be chosen for special
treatment.
The idea here is to select an energetic binary containing the massive
BH surrounded only by a small number of perturbers in the central density
cusp.
The latter requirement invariably implies a relatively small semi-major
axis, say $a_{\rm BH} \simeq R_{\rm cl}$, where
$R_{\rm cl} \simeq 10 r_{\rm h}/N$ is the adopted close encounter distance
for standard regularization treatment.
Upon selection of a suitable binary, a few nearby perturbers are added to
the new subsystem which is initialized as a composite particle for direct
$N$-body integration in the usual way (Aarseth 2003a).

Subsequently a variety of heuristic conditions have been employed to
control the subsystem size.
Particular attention is directed towards limiting the maximum membership
because of the $n (n - 1) /2$ force terms which are accumulated by the
accurate but time-consuming Bulirsch--Stoer (1966) integrator.
Thus the solution method is based on shared but variable time-steps and
each step requires many function evaluations for a tolerance of $10^{-12}$.
Likewise, internal members moving outside a specified distance are
removed from the subsystem and play the role as external perturbers while
tidal effects are important.
The processes of absorbing or emitting subsystem members entails careful
updating of the energy budget in order to maintain conservation at a
high level.

Particles outside the subsystem also play a role in driving the evolution.
External perturbers are selected for distances
$d < \left[2 m_j /m_0 \gamma_0\right]^{1/3} R_{\rm grav}$,
where $\gamma_0 = 10^{-6}$ is a small dimensionless parameter measuring
the relative perturbation at the boundary and $R_{\rm grav}$ is the
gravitational radius ($2 a_{\rm BH}$ for a binary).
Since the wheel-spoke formulation only yields regular equations for
interactions with the massive body, it is necessary to include a small
force softening in the singular terms, thereby permitting the smooth
treatment of near-collisions.
The softening length was first taken as $\epsilon = \epsilon_0 R_{\rm grav}$
with $\epsilon_0 \simeq 0.01$ and later reduced to $\epsilon_0 = 0.001$
without experiencing numerical problems.
It is updated every time the subsystem is reduced to a dominant binary
which occurs frequently.
Consistent potential energy corrections are then made at each change in
membership.
It should be emphasized that softening is only applied between the few
non-spoke internal interactions so that the essential dynamics is
maintained.

The subsystem solution is combined with the standard Hermite block-step
scheme (Makino 1991) which ensures synchronization and consistency.
Accordingly, the force on any subsystem perturber is evaluated carefully
by summation over all the internal members.
The adaptation to special-purpose GRAPE computers is based on the
standard {\ZZ {NBODY4}} code (Aarseth 2003a) where chain regularization
has been bypassed.
We note here that possible numerical problems for large mass ratios in
chain regularization are circumvented by the present Hamiltonian
formulation where the different relative motions are treated on an equal
footing.
Likewise, the absence of switching and re-initializing the chain for the
same membership is beneficial.

A special feature in the present investigation is the addition of
post-Newtonian terms in the relativistic regime.
This treatment necessitates the introduction of physical units.
Scaling to total energy $E = -1/4$ for the gravitational constant of
unity and $\sum m_i = 1$ gives an rms velocity of $2^{1/2}/2$ and velocity
unit $V^{\ast}$ in ${\rm km~s}^{-1}$ once the total mass and length scale
is specified.
Hence the speed of light is given by $c = 3 \times 10^5 / V^{\ast}$ in
model units.
The resulting equation of motion can be written in a convenient form by
(Blanchet \& Iyer 2003, Mora \& Will 2003)
\be
\frac {d^2 {\bf r}}{d t^2} \,=\,
 \frac {M}{r^2} \left[(-1 + A) \frac {\bf r}{r} + B {\bf v} \right ] \,,
\ee
where the scaled quantities $A$ and $B$ represent the accumulated
relativistic effects associated with the separation vector $\bf r$ and
relative velocity $\bf v$.
The latter is readily obtained from the momentum transformations when due
allowance is made for the non-zero momentum of $m_0$ in the local frame.

In the present formulation the perturbing {\it force} is required for
consistency with the equations of motion (cf. Mikkola \& Aarseth 1993).
Consequently, the desired expression for the dominant two-body motion takes
the form
\be
\begin{split}
 {\bf F}_{\rm GR} =
\frac {m_i m_0}{c^2 r^2} \left [ (A_1 + \frac {A_2}{c^2} +
 \frac {A_{5/2}}{c^3} + \frac {A_3}{c^4}) \frac {\bf {r}}{r} + \right. \\
 ~~~~~~~~~~~~~~~~~~~~~~~~~~\left. + (B_1 + \frac {B_2}{c^2} +
 \frac {B_{5/2}} {c^3} + \frac {B_3}{c^4}) \bf{v} \right ] \,.
\end{split}
\ee

In view of the increasing number of operations involved for the higher
orders, it is of interest to implement an efficiency scheme depending on
time-scale and order.
Such an experimental approach aims at providing a qualitative description
of the inspiralling process without any predictions for the final
coalescence event.

The classical time-scale for radiation energy loss is employed for
decision-making purposes.
In $N$-body units the instantaneous value, $a /{\dot a}$, for a large mass
ratio is given by (Peters 1964)
\be
\tau_{\rm GR} \,\simeq\, 
 \frac {5 a^4 c^5} {64 m_i m_0^2} \frac {(1 - e^2)^{7/2}}{g(e)} \,,
\ee
where $g(e)$ is a known function ($\simeq 4.5$ for $e = 1$).
Provided the time-scale falls below a specified value, the radiation terms
$A_{5/2}$ and $B_{5/2}$ are added to any Newtonian perturbations for the
dominant interaction.
Other terms are activated progressively on subsequent reduction of
$\tau_{\rm GR}$.
Moreover, comparisons with the full scheme for a nearly isolated binary with
circular or eccentric orbit yield final coalescence times in essential
agreement.
Energy conservation is monitored by separate integration of the
post-Newtonian perturbation according to
\be
 \Delta E_{GR} \,=\,
 \int {\bf F}_{\rm GR} \cdot {\bf v} \, d t \,,
\ee
converted to the appropriate regularized form (Mikkola \& Aarseth 1993).
Note that satisfactory conservation does not guarantee the solution here and
is merely a consistency check; hence comparison with known two-body solutions
should be carried out.

Since the present full formulation is applicable for quite large deviations
from Newtonian dynamics we define GR coalescence in the traditional way as
three Schwarzschild radii by
\be
 r_{\rm coal} \,=\, \frac {6 (m_i + m_0)} {c^2} \,,
\ee
or about $8 \times 10^{-10}$ in $N$-body units for the standard parameters.
This represents over five orders of magnitude in distance range with respect
to the initial close encounter length scale, $R_{\rm cl}$.

\section{INITIAL CONDITIONS}

Each simulation project requires careful considerations concerning the
initial conditions.
In the present work, the introduction of a massive central body may have a
significant effect on other nearby particles.
Since the emphasis is on studying the evolution of the innermost part,
it is desirable to start with an approximate equilibrium distribution.
Following an earlier formulation (Aarseth 2003b), we adopt a cusp-like
stellar density profile
\be
 \rho (r) \,=\, \frac {1}{r^{1/2} (1 + r^{5/2}) } \,.
\ee
Given a central body of mass $m_0$, the corresponding 1D velocity dispersion
is generated by (Zhao 1996)
\be
\sigma^2(r) = \frac{1}{\rho(r)} \int_{r}^{\infty}
\frac{\rho(r)}{r^2} [m(r) + m_0] dr \,,
\ee
where $m(r)$ is the enclosed mass within $r$.

Several models with $N = 1 \times 10^5$ equal-mass particles of
$1~{\rm M}_{\odot}$ are studied.
Based on the discussion above, we take the BH mass as $N^{1/2} {\bar m}$
which corresponds to $m_0 = 3 \times 10^{-3}$ in $N$-body units with
$m_i = 1 \times 10^{-5}$.
This mass ratio is relatively modest compared to typical values of
$m_{\rm BH} \simeq 0.01$ used in many investigations of massive binary
components.

\begin{figure}
\centering
\epsfig{file=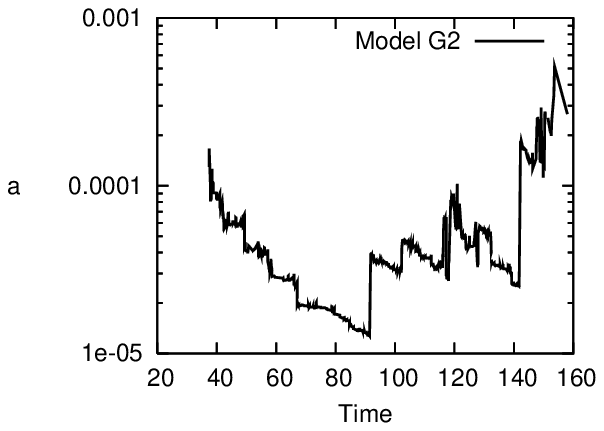, width=8cm, height=7cm}
\epsfig{file=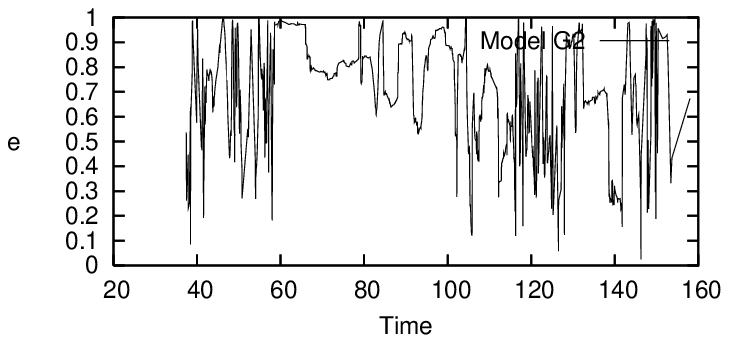, width=8cm, height=5cm}
\caption{Semi-major axis and eccentricity as functions of time, model G2.
GR coalescence occurred at $t = 92, 104, 117, 120, 130, 144, 152.$}
\label{fig1}
\end{figure}

The choice of total mass and half-mass radius determines the scaled speed
of light in the simulation and hence the influence of any post-Newtonian
interactions.
Since this investigation is a first attempt to evaluate the code performance,
standard star cluster parameters would appear to be outside the likely range
of interest unless exceptionally large eccentricities are reached.
For this reason we first adopt a more conservative length scale of 0.1~pc
which yields an initial half-mass radius $r_{\rm h} = 0.094$~pc.
With the velocity scaling $V^{\ast} \simeq 66$~${\rm km~s}^{-1}$, this makes
$c \simeq 4600$ in $N$-body units.
Consequently, an energetic binary with semi-major axis
$a \simeq 3 \times 10^{-5}$ would need an eccentricity exceeding 0.99 for
the radiation time-scale to fall below 400 $N$-body units which might be a
reasonable criterion for initiating the post-Newtonian treatment since
most simulation times are considerably shorter.
This compares to the smallest initial central binary of size
$a = 2.5 \times 10^{-4}$ in a typical model and only one more below
$1 \times 10^{-3}$.
Hence a significant evolution is required to achieve a factor of 10 shrinkage
of the most energetic binary, especially bearing in mind the increased
central velocity dispersion associated with the massive body.

The early cluster evolution exhibits small departures from overall
equilibrium as defined by the virial energy ratio.
Likewise, the number of particles inside the innermost fixed radii do not
show significant changes, in accordance with expectations for an equilibrium
model.

Several models assume that the interactions are between point-mass particles.
However, it is also of interest to study finite-size objects.
The possibility that stars are disrupted by the BH adds another
complication.
We introduce the tidal disruption distance (Magorrian \& Tremaine 1999),
\be
 r_t = \left ( \frac {m_0} {m_i} \right )^{1/3} r^{\ast} \,,
\ee
and adopt $r^{\ast} = 5 \times 10^{-5}$~au for white dwarfs of
$1~{\rm M}_{\odot}$.
In these models, any star inside this distance is removed from the
calculation and its mass added to the BH instantaneously.
The question of whether such orbits may be modified by post-Newtonian
effects before disruption depends on the cluster parameters.

\section{NUMERICAL RESULTS}

\begin{figure}
\centering
\epsfig{file=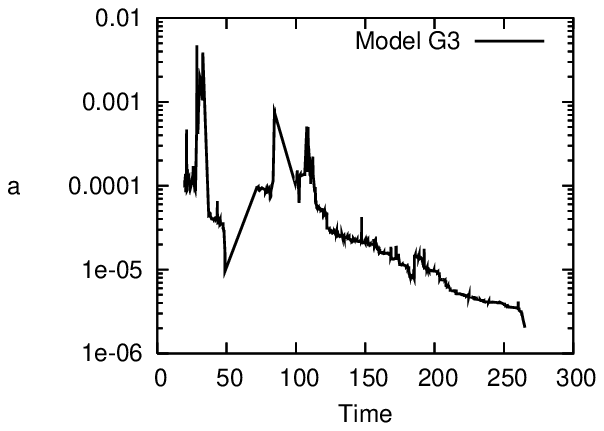, width=8cm, height=7cm}
\epsfig{file=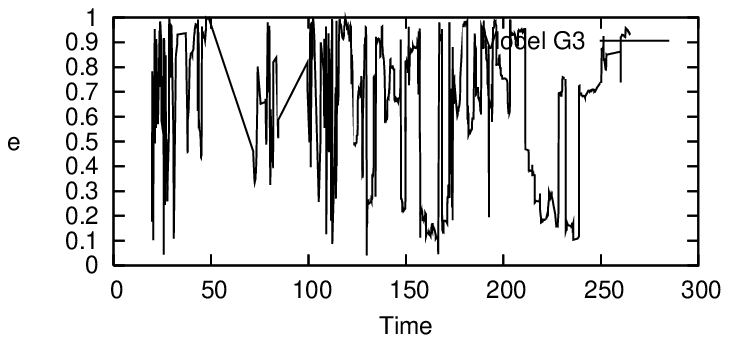, width=8cm, height=5cm}
\caption{Semi-major axis and eccentricity as functions of time, model G3.
Coalescence took place at $t = 22, 30, 50, 84, 104, 186.$}
\label{fig2}
\end{figure}

We first present some results for idealized models which have been
studied over at most a few hundred $N$-body time units and is
sufficient to illustrate the general behaviour.
Figure~1 displays the semi-major axis of the innermost bound orbit
in model G2 ($r_{\rm h} = 0.09$ pc), together with the corresponding
eccentricity.
Since there are seven GR coalescence events, the downwards trend of
the semi-major axis is replaced by the next most strongly bound orbit,
whereupon the shrinkage continues.
Inspection of the eccentricity graph reveals a number of spikes, with
both high and low values which are the hall-mark of Kozai cycles.
The growth in eccentricity preceding significant GR radiation loss
can be substantial, with pre-relativistic values exceeding
$e \simeq 0.99999$ on several occasions.
Likewise, the associated predicted maxima are often very close to
unity when the inclination is near $90^{\circ}$.
However, not all such configurations are sufficiently stable for
GR conditions to develop.

\begin{figure}
\centering
\epsfig{file=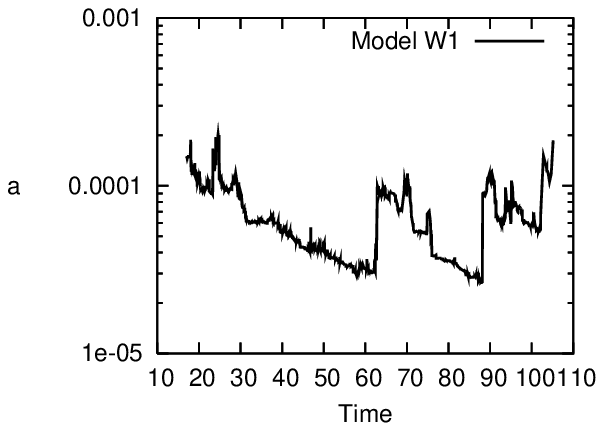, width=8cm, height=7cm}
\epsfig{file=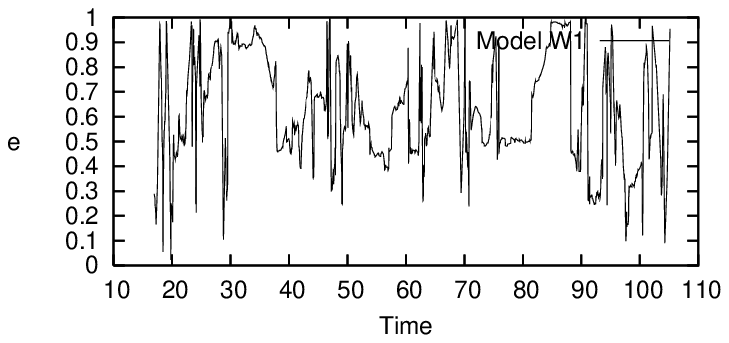, width=8cm, height=5cm}
\caption{Semi-major axis and eccentricity as functions of time, model W1.
Disruption occurred at $t = 25.0, 62.8, 70.2, 88.2, 95.2, 102.5, 104.8$.}
\label{fig3}
\end{figure}

The second model (G3), shown in Fig.~2, exhibits a different behaviour
of the smallest semi-major axis, this time with six inspiralling events.
Note that during two epochs ($t \simeq 50-70$ and $86-98$) the
central subsystem is too weakly bound for the special treatment.
After an irregular early phase there is a downward trend which is barely
affected by the last coalescence.
At termination the semi-major axis decreased below the last plotting
point to $1.2 \times 10^{-6}$ and would soon have satisfied the
coalescence criterion.
During the non-relativistic stage the innermost semi-major axis tends
to shrink as a consequence of dynamical evolution.
However, the shrinkage would progress much further but for the large
eccentricity which therefore places a lower limit on the period before
the GR effect intervenes.
Each coalescence is associated with a substantial loss of radiation
energy with an accumulated value $\Delta E_{\rm GR} \simeq -181$,
consistent with the final Newtonian binding energies.
This may be compared with a small systematic drift of
$-2 \times 10^{-5}$ units in the total energy which should be conserved.

These early models employ the point-mass assumption as in most current
work.
Based on the promising experience gained for point-mass particles it 
is worth while to enlarge the investigation and include finite-size
objects.
Given the uncertain status of neutron star populations in globular
clusters, it may be more appropriate to consider white dwarfs which
should be present with significant abundance in the cores.
Since the code can handle a general stellar distribution at a specified
age, a study of more realistic mass functions will be undertaken in due
course.

\begin{figure}
\centering
\epsfig{file=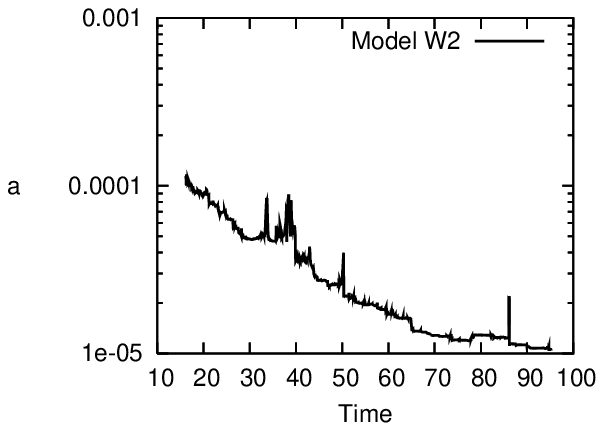, width=8cm, height=7cm}
\epsfig{file=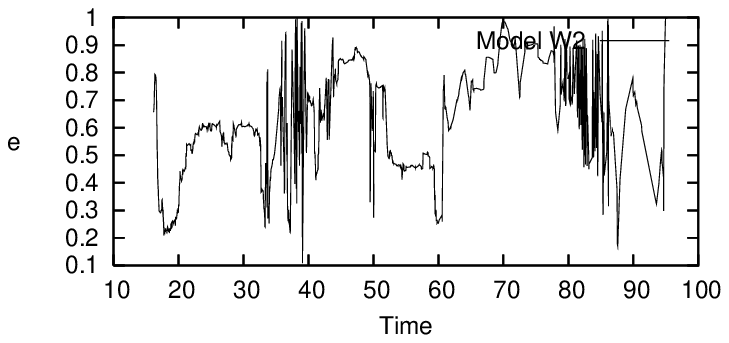, width=8cm, height=5cm}
\caption{Semi-major axis and eccentricity as functions of time, model W2.
In this model there was just one disruption at the end.}
\label{fig4}
\end{figure}

In the second set of models, the field stars are taken to be white
dwarfs with the characteristic radii and masses quoted above.
For continuity we begin with a similar cluster half-mass radius of
0.09~pc (model W1).
Figure 3 gives the smallest semi-major axis as a function of time,
together with the corresponding eccentricity.
There are seven disruptions in this interval, with the relevant times
quoted in the figure caption.
All but one of these events show significant GR energy loss while the
exception is a head-on collision with large pre-relativistic
eccentricity, $e = 0.999999$.
After the seventh disruption, the accumulated relativistic energy was
$\Delta E_{\rm GR} \simeq -0.31$ which exceeds the initial cluster
energy.
Once again the eccentricity plot reveals the characteristic spikes
associated with Kozai cycles.
Note that the corresponding time-scales are sometimes shorter than the
plotting intervals so that fine structure is often missing.

The cluster half-mass radius was also varied in order to explore the
length-scale dependence.
First, a large half-mass radius with $r_{\rm h} = 2.8$~pc was chosen
for model W2.
This gives a smaller disruption radius of
$r_{\rm t} \simeq 5 \times 10^{-10}$ compared to $1.6 \times 10^{-8}$
above.
The results shown in Fig.~4 contain just one disruption at a comparable
time to Fig.~3 ($t_{\rm f} \simeq 95$).
Again a Kozai cycle was apparent, with $e_{\rm max} \simeq 0.9998$.
A rather long period of circularization then took place before
termination ($e = 0.1, r = 5 \times 10^{-10}$).
Until this time, the previously accumulated relativistic energy loss
was small which indicates that no other critical approaches were
present.

\begin{figure}
\centering
\epsfig{file=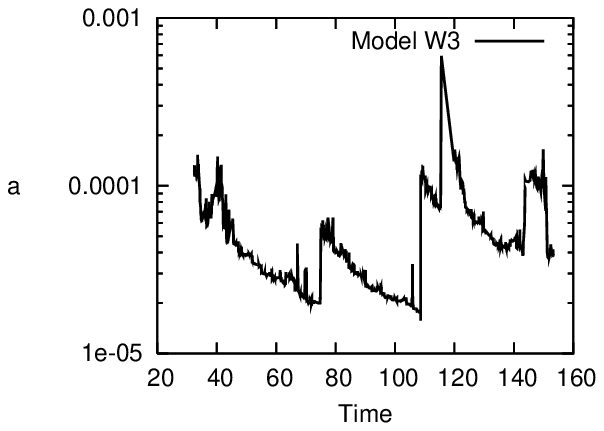, width=8cm, height=7cm}
\epsfig{file=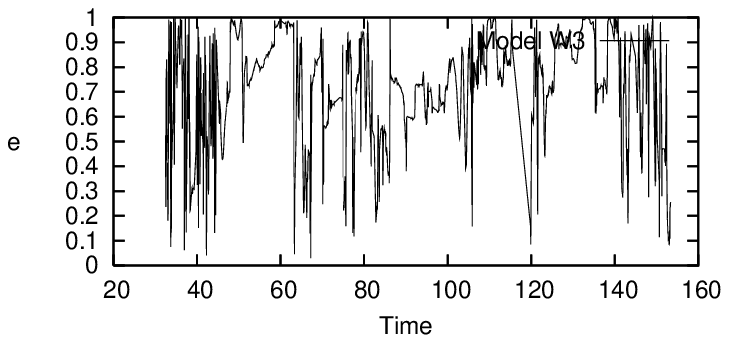, width=8cm, height=6cm}
\caption{Semi-major axis and eccentricity as functions of time, model W3.
Five disruption events occurred at $t = 38.7, 75.0, 108.6, 115.6, 143.8$.}
\label{fig5}
\end{figure}

In the final model W3 a cluster of intermediate half-mass radius
$r_{\rm h} = 0.9$~pc was studied.
A collision with small {\it predicted} pericentre for a wide orbit
($a \simeq 0.02, \,1 - e = 10^{-7}$) occurred during the early stage
before the compact subsystem formation.
All the five disruption events exhibited maximum eccentricities
exceeding 0.9999 shortly before termination, evaluated at a
non-relativistic distance $r \simeq 10^{-5}$.
Moreover, in each case the energy loss by equation~(8) yielded an amount
$\Delta E_{\rm GR} \simeq -0.5$.
This shrinkage corresponds to a semi-major axis
$a_{\rm GR} \simeq 3 \times 10^{-8}$ which agrees with the Newtonian value
obtained just before the final plunge.

The density distribution in the central region is also of interest.
At typical advanced stages of inner binary evolution
($a \simeq 2 \times 10^{-5}$), there are only about 10 members inside a
distance of $300 a$.
Moreover, the second innermost semi-major axis tends to be well separated
from the innermost one, thereby giving opportunities for favourable Kozai
cycles.
Only a few of the dozen or so most extreme eccentricity excursions result
in significant energy loss and may therefore be considered rare events.

\section{POST-NEWTONIAN ASPECTS}

The results presented above show examples of relativistic effects
leading to coalescence or disruption.
Although the inner binary shrinkage is often considerable, the eccentricity
plays a key role in initiating the relativistic stage.
The process by which large eccentricities are attained is mainly due to
third-body secular perturbations known as Kozai (1962) cycles.
This requires the perturber to remain in a long-lived orbit around the
innermost binary with initial inclination exceeding about $40^{\circ}$.
Conservation of angular momentum yields a relation connecting the
inclination and inner eccentricity in the form
$cos^2 \theta (1 - e^2) = {\rm const.}$
The period for the induced inner eccentricity variation is given by
\be
 T_{\rm Kozai} = \frac {T_{\rm out}^2} {T_{\rm in}}
\left( \frac {1 + q_{\rm out}} {q_{\rm out}}\right) (1 - e_{\rm out}^2)^{3/2}
 f(e_{\rm in}, \omega_{\rm in}, \psi) \,,
\ee
where $q_{\rm out} = m_2/(m_1 + m_0)$ is the mass ratio for the outer orbit
with elements $a_{\rm out}, \, e_{\rm out}$ and the function $f$ which
depends on orbital elements is usually of order unity (Heggie 1996).

An expression is also available for the maximum eccentricity of the inner
orbit, $e_{\rm max}$, which is of considerable interest (Heggie 1996).
Since the period ratio $T_{\rm out} / T_{\rm in}$ may be taken of order
10 for a long-lived system here, this gives
$T_{\rm Kozai} \simeq 3000 T_{\rm out} (1 - e^2_{\rm out})^{3/2}$.
Hence a Kozai period of about $1000\, T_{\rm out}$ may suffice for
$e_{\rm out} \simeq 0.7$ and a moderately high inclination, which is
frequently seen in such simulations.

A new decision-making scheme has been developed to introduce the different
post-Newtonian terms for separate time-scales.
Although these terms are increasing in complexity, this is done only partly
for computational efficiency.
The idea of considering the additional perturbations progressively
(Aarseth 2003b) has also been tested by a stand-alone code for the three-body
problem\footnote{See the public three-body regularization code
{\tt TRIPLE3} at {\tt http://www.ast.cam.ac.uk/$\sim$sverre/multireg.}}
(Aarseth \& Zare 1974).
In the present work, the emphasis is on achieving a qualitatively correct
description of the evolution towards coalescence or tidal disruption
without aiming for detailed predictions.

Since the main decision-making is based on the two-body elements $a$ and $e$,
it is desirable to use relativistic definitions when appropriate.
This procedure was not adopted for most of the work reported here.
However, the elements are usually evaluated {\it outside} the semi-major
axis of predominantly highly eccentric orbits where GR effects are
negligible.
For a given level of post-Newtonian implementation, we add the relevant
contributions from the two first $A$ and $B$ terms of equation~(6)
according to the relativistic expansion (Mora \& Will 2003).
Without these corrections, the classical elements show a range of values
depending on orbital phase and $v/c$.
Now the modified two-body energy and angular momentum yield consistent
values of the osculating semi-major axis and eccentricity, as can be
verified by the available three-body code for an isolated binary.
We remark that the relativistic contributions for the white dwarf models
are fairly modest, with typical values $v/c \simeq 4 \times 10^{-4}$ for
$r > a$ near disruption in the case $r_{\rm h} = 0.9$~pc.

In the case of a binary we define appropriate time-scales for activating the
perturbations in terms of the instantaneous radiation time-scale,
$\tau_{\rm GR}$.
Thus the radiation term 2.5PN is already included for $\tau_{\rm GR} < 1000$
$N$-body units.
The subsequent terms 1PN, 2PN and 3PN are then activated below the
experimental values of 100, 50 and 10, respectively.
Likewise, the relevant terms are excluded for increasing time-scales up to
$\tau_{\rm GR} > 2000$ for the GR radiation, with slightly more generous
limits being applied in order to prevent repeated switching.
Three-body tests show that the final coalescence times for close binaries of
different eccentricity are in essential agreement with the full scheme when
criteria of this type are employed for a range of parameters.
Regarding the choice of boundary values for different levels, the nominal
time-scale exceeds the actual coalescence time for circular orbits by a
small factor depending on $c$ due to the loss of angular momentum.
%However, the predicted time-scale {\it increases} for eccentric orbits
%during the final approach if small values are reached before termination.
At some advanced stage, it may be justified to adopt the unperturbed
two-body approximation (Peters 1964) or even define premature termination
in order to speed up the calculation.

The possible presence of Kozai cycles adds a further complication since
the precession terms act to de-tune the resonance.
In particular, the classical ``Mercury'' precession gives rise to a
pericentre advance in radians per orbit,

\be
\Delta \omega = \frac {6 \pi m_0 } {c^2 a (1 - e^2) } \,,
\ee
which may be substantial for some conditions of interest.
Unless existing post-Newtonian perturbations are present due to a
sufficiently short radiation time-scale, the first and second-order
precessions are introduced for $T_{\rm Kozai} < 10$.
Thus for an eccentric binary with $e = 0.99$ the precession condition gives
$\Delta \omega \simeq 0.01$ for a typical inner binary of size
$a = 1 \times 10^{-5}$ which is usually covered by either of these criteria.
More extreme stages may give rise to a faster advance which is reflected
in a shortening of the time-steps.
The basic three-body code, which contains identical post-Newtonian
expressions to the full code, was used to compare the pericentre advance
in the case of strong relativistic effects while suppressing the higher
order terms.
Here we employed an algorithm measuring the angle between successive
apocentre turning points which gave good agreement with equation~(14)
and therefore provides an independent check on the first-order solution.

Two small bodies orbiting a much heavier one in relative isolation may be
sufficiently stable for Kozai cycles to be induced.
All that is required is that an outer orbit increases its energy and has
a favourable inclination for large eccentricity growth.
As has been noted above, this process is seen at later stages when the
inner core has been partially depleted.
Hence it appears that the final orbital shrinkage before relativistic
effects become important is closely correlated with the Kozai resonance.
However, the process by which the second most energetic binary migrates
inwards in the excavated core and sometimes acquires a large inclination
(near $90^{\circ}$) needs to be investigated further.

\begin{figure}
\centering
\epsfig{file=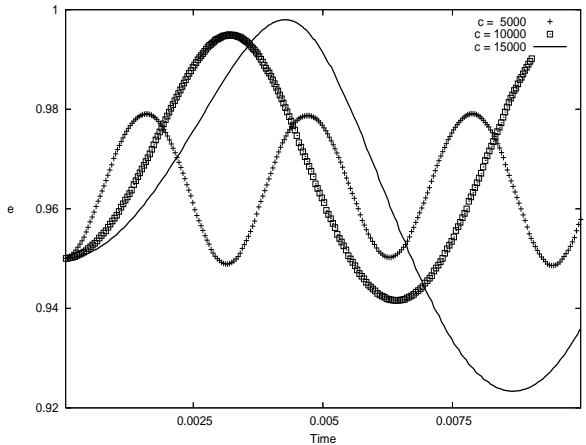, width=8cm, height=6cm}
\caption{Eccentricity as function of time for three values of $c$.
The elements of the hierarchical three-body configuration are defined
in the text.}
\label{fig6}
\end{figure}

Idealized Kozai cycles may be studied by the three-body
regularization code which employs the same GR terms.
Figure 6 illustrates the behaviour of the inner eccentricity for a
long-lived hierarchical triple with initial inclination near
$90^{\circ}$ using $c = 5\,000, 10\,000, \,15\,000$ to represent
different mass or length scales.
Simulation mass units are adopted with
$a_0 = 1 \times 10^{-5},\,a_1 = 4.5 \times 10^{-5}$
and eccentricities $e_0 = 0.95,\,e_{\rm out} = 0.2$.
From these parameters, $e_{\rm max} = 0.9999$ while the peak at
0.9980 is reached for $c = 15\,000$, with $c$ similar to model W3.
This may be sufficient for GR shrinkage to be achieved on a longer
time-scale.
In this example, the peak eccentricity is maintained over $\simeq 100$
Kozai cycles with pericentre advance
$\Delta \omega \simeq 0.006$ radians per orbit, and the slope
of the radiation energy loss corresponds to an inspiralling time
$a/{\dot a} \simeq 100$.
The basic Kozai period is a factor of 5 shorter than given by
equation~(13).
Comparison with an earlier analysis (Miller \& Hamilton 2002, eq. 6)
may also be of interest.
We obtain $\theta_{\rm PN} \simeq 0.29$ for the dimensionless
post-Newtonian Hamiltonian in the realistic example.
Although consistent, the expression $e_{\rm max}=1-\theta_{\rm PN}^2/9$
applies to the restricted three-body problem.

In a second example,
$a_0 = 5 \times 10^{-5}, \,a_1 = 2.3 \times 10^{-4}$, with $c = 15\,000$
and inclination $100^{\circ}$.
Here the inner semi-major axis shrank to $\simeq 1 \times 10^{-5}$ after
some $3 \times 10^4$ outer periods.
During the early stage, up to $a_0 \simeq 3 \times 10^{-5}$, the
eccentricity was maintained near 0.999 as the result of competition
betwen the Kozai cycle and relativistic decay.
This was eventually superseded by pure decay consistent with the
analytical expression for $\dot e$ (Peters 1964).

The case of hyperbolic encounters with the BH has been considered in a
few simulations only.
For this purpose we introduce the capture radius (Quinlan \& Shapiro 1989),
\be
 r_{\rm cap} =
b \left [ \frac {m_0 m_i (m_0+m_i)^{3/2} }{c^5 v_{\infty}^2} \right ]^{2/7}\,,
\ee
where b = 2.68 and $v_{\infty}$ is the velocity at infinity in $N$-body
units.
With the present parameters in the standard white dwarf model we obtain
$r_{\rm cap} \simeq 9 \times 10^{-9}$ for a typical excess velocity in the
core $v_{\infty} = 1$ and hence capture due to gravitational radiation is
not likely to occur before disruption.
On the other hand, the capture process is quite feasible in the point-mass
case.

\section{DISCUSSION}

The work presented in this paper should be considered exploratory.
Although the post-Newtonian formulation itself is well known, a practical
implementation presents many additional problems.
The attempt to introduce a scheme based on different time-scales without
sacrificing essential dynamics appears to have been successful.
Likewise, the numerical challenges facing the new implementation have
been resolved.
Thus it has been demonstrated that the GRAPE-6 combined with the
wheel-spoke regularization code is suitable for studying stellar systems
with $10^5$ members and an intermediate massive single BH.
A closer analysis of the performance shows that the overheads connected
with the central subsystem represent less than 10 per cent (for 64-bit
operating system on GRAPE-6A) which is a small price for an accurate
treatment.
The surprisingly short time-scale for interesting developments also
means that significant results can be obtained in a few days even with
the smaller GRAPE-6A.

One advantage of using a multiple regularization method, as opposed to
the two-body formulation of {\ZZ {NBODY6}} (Aarseth 2003a), is that the
complicated derivatives of the GR terms are not required
(cf.~Kupi et al. 2006).
On the other hand, it appears that the slight defect of introducing
a small softening between the low-mass members of the subsystem is
justified\footnote{A scaled softening $\epsilon_0 = 0.001$ and
$R_{\rm grav} \simeq 10^{-5}$ would correspond to a white dwarf radius
in model W3.} (and even smaller values may be acceptable).
So far, the 3PN terms have been added for the shortest time-scales,
$\tau_{\rm GR}$.
This was done for illustrative purposes and may not be required since
the 1PN and 2PN terms act to de-tune the Kozai resonance.
Contributions from the 3.5PN terms would in principle give rise to a
recoil velocity; however, this is likely to be very small for the
present mass ratio.

So far tidal capture and oblateness effects have not been included
(Mardling \& Aarseth 2001).
General considerations suggest that this process would tend to increase
the supply of stars into the loss-cone.
On the technical side, the implementation would be more complicated.
Thus with two-body regularization it is convenient to apply the
orbital modifications as an impulse at pericentre while in chain
regularization this procedure requires more care (Aarseth 2003a).
Likewise, hyperbolic capture of neutron stars or point-mass particles
by gravitational gravitation (cf. equation~(15)) would require a similar
treatment and may be worth the effort.

Issues connected with a larger length-scale should also be addressed.
From equation~(7) and the definition of $c$ we have
$\tau_{\rm GR} \propto r_{\rm h}^{5/2}$ in $N$-body units for the same
semi-major axis and eccentricity.
Hence the two-body elements need to be more favourable in order for the
GR regime to be reached with a comparable effort.
Increasing the half-mass radii of the white dwarf cluster models
indicates that relativistic effects may still be appreciable.
Further explorations of such models are therefore desirable.

The intriguing question of the relative importance of resonant
relaxation (Rauch \& Tremaine 1996) versus Kozai cycles needs to be
investigated.
Other authors have pointed to the connection (Hopman \& Alexander 2006)
and, as reported above, there is no doubt that the latter process plays
a crucial role triggering the early energy loss.
Typical examples have been identified showing extreme eccentricity
evolution for long-lived configurations with semi-major axis ratio of
about 10.
Even so, the actual duration may well be too short to be resolved by
the present plotting procedure in some cases.
As for de-tuning of the Kozai cycle, characteristic values
$\Delta \omega \simeq 0.05$ are obtained from equation~(14) for the
first-order precession during critical stages involving white dwarfs
and this may not be excessive.

The importance of Kozai cycles in $N$-body simulations has been
emphasized before (Iwasawa et al. 2006).
In this work, the interaction of three massive objects
($m_{\rm BH} = 0.01$) in a system of $N \simeq 10^5$ low-mass particles
included the gravitational radiation term with fairly small values of $c$.
The formation of hierarchical triples involving the three massive bodies
frequently led to the ejection of one member by the sling-shot mechanism
or, alternatively, to coalescence of the inner binary.
The corresponding eccentricity displayed large and small spikes which
is characteristic of Kozai cycles.
As demonstrated above, including the GR precession does not suppress
the eccentricity spikes.
Given the enhanced precession rates during the approach to GR conditions,
more work is needed to clarify this behaviour.
In this connection, we note that the predicted maximum eccentricity
is often reached before triggering the final stages leading to
coalescence or tidal disruption.

It is interesting to compare the numerical challenges of studying one
or two massive objects in a stellar system.
In the latter case, the binary acts to clear the inner region by ejecting
stars.
The depletion of short periods therefore makes for an easier technical
problem, provided a reliable method is available to deal with the
%massive binary\footnote{The so-called {\ZZ {TTL}} method already includes
%three post-Newtonian terms.}(cf. Mikkola \& Aarseth 2002, Aarseth 2003b).
massive binary Mikkola \& Aarseth 2002, Aarseth 2003b).
In either case quite large eccentricities are required for GR effects
to manifest themselves.
As far as the study of a single BH object is concerned, the wheel-spoke
formulation provides a practical scheme for including post-Newtonian
terms in a direct $N$-body code.

Applications to a star cluster model containing white dwarfs also show
that significant orbital shrinkage by energy radiation loss is possible
before disruption.
Moreover, high eccentricities by the Kozai mechanism can still be
achieved.
These processes lead to an enhanced rate of swallowing by the BH but
more extensive simulations are needed before the growth rate can be
determined.
Hence the present software development provides a powerful tool for
exploring both idealized and realistic problems of current interest.
It remains to be seen whether such systems would bear any imprints of
the swallowing process in the form of supernova events caused by
white dwarf detonation (Dearborn, Wilson \& Mathews 2005).
Finally, the purpose of the present work is to present a viable new
method, together with some results illustrating its capability.
Hopefully, future work will address issues relating to time-scales
and model dependence.

\section{ACKNOWLEDGEMENTS}

Much of the code refinements were done during a one month visit to
the National Astronomical Observatory, Japan, when a draft was also
produced.
The author would like to thank Professor Eiichiro Kokubo for his kind
hospitality.
This visit was sponsored by the ``GRAPE-DR Project'' grant provided
by MEXT, Japan.
It is a pleasure to thank Seppo Mikkola for technical assistance.
Discussions with Pau Amaro-Seoane and Rosemary Mardling clarified my
understanding of GR implementations and few-body dynamics, respectively.

\label{lastpage}


\begin{thebibliography}{99}

\bibitem[\protect\citeauthoryear{Aarseth}{2003a}]{aa03}
Aarseth S.J., Gravitational N-Body Simulations, Cambridge
Univ. Press, Cambridge

\bibitem[\protect\citeauthoryear{Aarseth}{2003b}]{ab03}
Aarseth S.J., 2003b, Ap\&SS,~285, 367

\bibitem[\protect\citeauthoryear{Aarseth \& Zare}{1974}]{az74}
Aarseth S.J., Zare K., 1974, Cel.~Mech.,~10, 185

\bibitem[\protect\citeauthoryear{Alexander}{1986}]{al86}
Alexander M.E., 1986, J.~Comp.~Phys.,~64, 195

\bibitem[\protect\citeauthoryear{Amaro-Seoane \& Freitag}{2006}]{af06}
Amaro-Seoane P., Freitag M., 2006, MNRAS, xxx, yyy
%Amaro-Seoane P., Freitag M., 2006, MNRAS, in press (astro-ph/0610478)

\bibitem[\protect\citeauthoryear{Baumgardt et al}{2006}]{ba06}
Baumgardt H., Hopman C., Portegies Zwart S., Makino J., 2006,
MNRAS, 372, 467

\bibitem[\protect\citeauthoryear{Baumgardt, Makino \& Ebisuzaki}{2004}]{bme4}
Baumgardt H., Makino J., Ebisuzaki T., 2004, ApJ,~613, 1133

\bibitem[\protect\citeauthoryear{Berczik et al}{2006}]{be06}
Berczik P., Merritt D., Spurzem R., Bischof H., 2006, ApJ,~642, L21

\bibitem[\protect\citeauthoryear{Blanchet \& Iyer}{2003}]{bi03}
Blanchet L., Iyer B., 2003, Class.~Quant.~Grav.,~20, 755

\bibitem[\protect\citeauthoryear{Bulirsch \& Stoer}{1966}]{bs66}
Bulirsch R., Stoer J., 1966, Num.~Math.,~8, 1

\bibitem[\protect\citeauthoryear{Dearborn, Wilson \& Mathews}{2005}]{dwm5}
Dearborn D.S.P., Wilson J.R., Mathews G.J., 2005, ApJ,~630, 309

\bibitem[\protect\citeauthoryear{Heggie}{1996}]{he96}
Heggie D.C., 1996, private communication

\bibitem[\protect\citeauthoryear{Hopman \& Alexander}{2006}]{ha06}
Hopman C., Alexander T., 2006, ApJ,~645, 1152

\bibitem[\protect\citeauthoryear{Iwasawa, Funato \& Makino}{2006}]{ifm6}
Iwasawa M., Funato Y., Makino J., 2006, ApJ,~651, 1059

\bibitem[\protect\citeauthoryear{Kozai}{1962}]{ko62}
Kozai Y., 1962, AJ,~67, 591

\bibitem[\protect\citeauthoryear{Kupi, Amaro-Seoane \& Spurzem}{2006}]{kas6}
Kupi G., Amaro-Seoane P., Spurzem R., 2006, MNRAS,~371, L45

\bibitem[\protect\citeauthoryear{Lee}{1993}]{le93}
Lee M.H., 1993, ApJ,~418, 147

\bibitem[\protect\citeauthoryear{Magorrian \& Tremaine}{1999}]{mt99}
Magorrian J., Tremaine S., 1999, MNRAS,~309, 447

\bibitem[\protect\citeauthoryear{Makino}{1991}]{ma91}
Makino J., 1991, ApJ,~369, 200

\bibitem[\protect\citeauthoryear{Mardling \& Aarseth}{2001}]{ma01}
Mardling R.A., Aarseth S.J., 2001, MNRAS,~321, 398

\bibitem[\protect\citeauthoryear{Mikkola \& Aarseth}{1993}]{ma93}
Mikkola S., Aarseth S.J., 1993, Celes.~Mech.~Dyn.~Ast.,~57, 439

\bibitem[\protect\citeauthoryear{Mikkola \& Aarseth}{2002}]{ma02}
Mikkola S., Aarseth S.J., 2002, Celes.~Mech.~Dyn.~Ast.,~84, 343

\bibitem[\protect\citeauthoryear{Milosavlevi\'c \& Merritt}{2001}]{mm01}
Milosavlevi\'c M., Merritt D., 2001, ApJ,~563, 34

\bibitem[\protect\citeauthoryear{Mora \& Will}{2003}]{mw03}
Mora T., Will C., 2003, gr-qc/0312082

\bibitem[\protect\citeauthoryear{Peters}{1964}]{pe64}
Peters P.C., 1964, Phys.~Rev.,~136, B1222

\bibitem[\protect\citeauthoryear{Preto, Merritt \& Spurzem}{2004}]{pms4}
Preto M., Merritt D., Spurzem R., 2004, ApJ,~613, L109

\bibitem[\protect\citeauthoryear{Quinlan \& Shapiro}{1989}]{qs89}
Quinlan G.D., Shapiro S.L., 1989, ApJ,~343, 725

\bibitem[\protect\citeauthoryear{Rauch \& Tremaine}{1996}]{rt96}
Rauch K.P., Tremaine S., 1996, New~Astron.,~1, 149

\bibitem[\protect\citeauthoryear{Szell, Merritt \& Mikkola}{2005}]{smm5}
Szell A., Merritt D., Mikkola S., 2005, in Buchler, J.R.,
Gottesman S.T, Mahon M.E. eds, Nonlinear Dynamics in
Astronomy and Physics, Ann.~New York Acad.~Sci.,~1045, 225

\bibitem[\protect\citeauthoryear{Zare}{1974}]{za74}
Zare K., 1974, Celes.~Mech.,~10, 207

\bibitem[\protect\citeauthoryear{zhao}{1996}]{zh96}
Zhao H., 1996, MNRAS,~278, 488

\end{thebibliography}
\end{document}